\documentstyle[sprocl]{article}

\input{epsf}
\def\Journal#1#2#3#4{{#1} {\bf #2}, #3 (#4)}

\def\NPB{{\em Nucl. Phys.} B}
\def\PLB{{\em Phys. Lett.}  B}
\def\PRL{\em Phys. Rev. Lett.}
\def\PRD{{\em Phys. Rev.} D}
\def\ZPC{{\em Z. Phys.} C}

\def\be{\begin{equation}}
\def\ee{\end{equation}}
\def\bea{\begin{eqnarray}}
\def\eea{\end{eqnarray}}

\begin{document}

\title{HIGH $k_T$ MESON PRODUCTION: A DIFFERENT WAY TO PROBE HADRON
STRUCTURE\footnote
{Invited talk given at the Electron Polarized Ion Collider Workshop
(EPIC99), at the Indiana University Cyclotron Facility, Bloomington,
Indiana, USA, 8--11 April 1999} }

\author{CARL E. CARLSON}

\address{Nuclear and Particle Theory Group, Physics Department,\\
College of William and Mary, Williamsburg, VA  23187-8795 }

\vglue -1.7 cm  
\hfill WM-99-111; hep-ph/9905492
\vglue .7 cm


\maketitle\abstracts{
Hard, or high transverse momentum, pion  photoproduction can be a tool
for probing the parton structure of the beam and target. We discuss the
perturbative and soft processes that contribute, and show how regions
where perturbative processes dominate can give us the parton structure
information.  Polarized initial states are needed to get information
on polarization distributions.  Current polarization asymmetry data
is mostly in the soft region.  However, with the proposed
EPIC machine parameters, determining the polarized gluon distribution
using hard pion photoproduction appears quite feasible.}

\section{Semi-Exclusive Processes as Probes of Hadron Structure}

Information about hadron structure, in the form of distribution
functions or quark wave functions, classically comes from deep
inelastic scattering, Drell-Yan processes, or coincident
electroproduction~\cite{flavor}. What we shall study here is the
possibility of getting the same kind of information from
photoproduction of hard, which means high transverse momentum, pions or
non-coincident electroproduction of the
same~\cite{many,peralta,cw93,acw97,acw98,bdhp99}.  The production can
proceed by a number of processes, including direct pion production,
direct photon interactions followed by parton fragmentation, resolved
photon processes, and soft processes.  The first of these, direct pion
production can also be called short-distance or isolated pion
production. In the following, we will discuss how the first two of these
processes can give hadronic distribution function and wave function
information.  Information can also come from resolved photon
processes in the right circumstances, but we shall not pursue them
here. Soft processes are from the present viewpoint an annoyance, but
one we need to discuss and estimate the size of.  All the processes
will be defined and discussed below.

To begin being more explicit, the semi-exclusive process we will discuss
is 

\begin{equation}
\gamma + A \rightarrow M + X  ,
\end{equation}
where $A$ is the target and $M$ is a meson, here the pion.  The process
is perturbative because of the high transverse momentum of the pion,
not because of the high $Q^2$ of the photon.  Our considerations will
also apply to electroproduction,

\begin{equation}
e+ A \rightarrow M + X
\end{equation}
if the final electron is not seen.  In such a case, the exchanged
photon is nearly on shell, and we use the Weiz\"acker-Williams
equivalent photon approximation~\cite{bkt71} to relate the electron and
photon cross sections,

\begin{equation}
d\sigma(eA \rightarrow MX) = \int dE_\gamma \, N(E_\gamma) 
      d\sigma(\gamma A \rightarrow MX),
\end{equation}
where the number distribution of photons accompanying the electron is a
well known function.  

In the following section, we will describe the subprocesses that
contribute to hard pion production, and in the subsequent section
display some results.  Section~\ref{summary} will be a short summary.


\section{The Subprocesses}


\subsection {At the Highest $k_T$}

\begin{figure}

\vskip 5mm

\centerline { \epsfxsize 3.5in \epsfbox{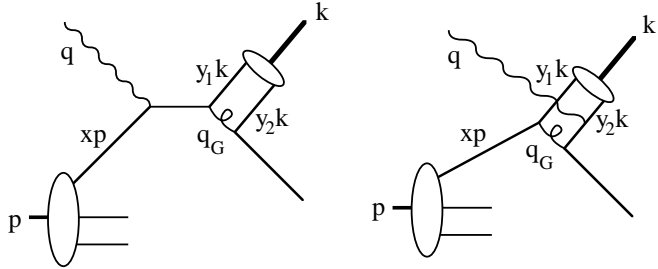}    }

\caption{Direct pion production.  The pion is produced in a  short
distance perturbatively calculated process, not by fragmentation of an
outgoing parton.  Thus this pion is kinematically isolated, and not
part of a jet.  Direct pion production gets important at very high
transverse momentum because the pion does not have to share the
available momentum with any other particles going in its direction. } 
\label{direct}

\end{figure}

At he highest possible transverse momenta, observed pions are directly
produced at short range via a perturbative QCD (pQCD) calculable
process~\cite{cw93,acw97,acw98,bdhp99}.  Two out of four lowest order
diagrams are shown Fig.~\ref{direct}.  The pion produced this way is
kinematically isolated rather than part of a jet, and may be seen
either by making  an isolated pion cut or by having some faith in the
calculation and going to a kinematic region where this process
dominates the others.  Although this process is higher twist, at the
highest transverse momenta its cross section falls less quickly than
that of the competition, and we will show plots indicating the
kinematics where it can be observed.


The subprocess cross  section for direct or short-distance pion
production is 

\begin{equation}
{d\hat\sigma \over dt} (\gamma q \rightarrow \pi^\pm q')
  = {128 \pi^2 \alpha \alpha_s^2 \over 27 (-t) \hat s^2}  I_\pi^2
   \left( {e_q \over \hat s} + {e_q' \over \hat u} \right)
 \left[  \hat s^2 + \hat u^2 + \lambda h (\hat s^2 - \hat u^2) \right] ,
\end{equation}
where $\hat s$, $\hat t = t$, and $\hat u$ are the subprocess
Mandlestam variables;  $\lambda$ and $h$ are the helicities of the
photon and target quark, respectively; and $I_\pi$ is an integral
related to the pion wave function at the origin in coordinate space,

\begin{equation}
I_\pi = \int {dy_1 \over y_1} \phi_\pi (y_1, \mu^2)  .
\end{equation}

\noindent In the last equation, $\phi_\pi$ is the distribution
amplitude of the pion, and describes the quark-antiquark part of the
pion as a parallel moving pair with momentum fractions $y_i$.  It is
normalized through the rate for $\pi^\pm \rightarrow \mu \nu$, and for
example,

\begin{equation}
\phi_\pi = {f_\pi \over 2 \sqrt{3}} 6 y_1 (1-y_1)
\end{equation}
for the distribution amplitude called ``asymptotic'' and for 
$f_\pi \approx 93$ MeV. Overall, of course, 

\begin{equation}
{d\sigma \over dx \, dt}(\gamma A \rightarrow \pi   X)
  = \sum_q  G_{q/A}(x,\mu^2) 
         {d\hat\sigma \over dt} (\gamma q \rightarrow \pi^\pm q')  ,
\end{equation}
where $G_{q/A}(x,\mu^2)$ is the number distribution for quarks of
flavor $q$ in target $A$ with momentum fraction $x$ at renormalization
scale $\mu$.

Let us note a number of points about direct pion production.

$\bullet$  For the photoproduction case, at least, the momentum
fraction of the struck quark is determined from experimental
observables.  This is like the situation in deep inelastic scattering,
where the experimenter can measure $x \equiv Q^2/2m_N \nu$ and the
theorist can prove that this $x$ is the same as the momentum fraction
of the struck quark, for high $Q$ and $\nu$.  For the present case,
define the momenta 

\begin{equation}
\gamma(q) + A(p) \rightarrow \pi(k) + X,
\end{equation}
and then the Mandlestam variables for the overall process,

\begin{equation}
s = (p+q)^2; \qquad t = (q-k)^2; \quad {\rm and} \quad u = (p-k)^2.
\end{equation}
Each of the Mandlestam variables is an observable, and the ratio

\begin{equation}
x = {-t \over s+u}
\end{equation}
is the momentum fraction of the struck quark.  We will let the reader
prove this.

$\bullet$ The gluon involved in direct pion production is well off
shell.  We will illustrate this by comparison to the pion
electromagnetic form factor.  For the gluon in
Fig.~\ref{direct}(right), 

\begin{equation}
q_G^2 = (xp-y_1k)^2 = y_1 x u  .
\end{equation}
(The gluon is Fig.~\ref{direct}(left) is farther off shell.) To get a
number, take
$E_\gamma = 100$ GeV and
$90^\circ$ in the center of mass.  Then $u \approx -95$ GeV$^2$, and
using 
$\langle y_1 \rangle \approx 1/3$ and $x \approx 1/2$ gives

\begin{equation}
\langle q_G^2 \rangle \approx -15 {\rm \ GeV\ }^2  .
\end{equation}

In a calculation of $F_\pi$,   

\begin{equation}
\langle q_G^2 \rangle = \langle y_1 y_2' \rangle q^2 \approx 
     {1\over 9} q^2,
\end{equation}
where $y_1$ and $y_2'$ are the momentum fractions, in the incoming and
outgoing pion, of the quark that does not absorb the photon.  Hence to
match the above direct pion production kinematics requires measuring
$F_\pi$ at a momentum transfer
$|q^2| = 135$ GeV$^2$.  We come much closer to asymptopia in direct pion
production than in a thinkable pion form factor measurement!

$\bullet$  Without polarization, we can measure $I_\pi$, given enough
trust in the other parts of the calculation.  This $I_\pi$ is precisely
the same as the $I_\pi$ in both $\gamma^* \gamma \rightarrow \pi^0$
(which is measured in $ee \rightarrow ee\pi^0$) and 
$e\pi^\pm \rightarrow e \pi^\pm$ (which gives the pion form factor
$F_\pi$).  Presently, the experimental results for 
$\gamma^* \gamma \rightarrow \pi^0$ agree with the theoretical results
using the asymptotic distribution amplitude mentioned earlier, but the
results for $F_\pi$ disagree with the same.  Thus there is room for a
third process in measuring $I_\pi$.

$\bullet$ We also have polarization sensitivity in direct pion
production.  For $\pi^+$ production at high $x$,

\begin{equation}
A_{LL} \equiv {        \sigma_{R+} - \sigma_{L+} 
                                \over 
                       \sigma_{R+} + \sigma_{L+}          }
   =  {s^2 - u^2 \over s^2 + u^2  }  {\Delta u(x) \over u(x) }
\end{equation}
where $R$ and $L$ refer to the polarization of the photon, and $+$
refers to the target, say a proton, polarization.  Also, inside a $+$
helicity proton the quarks could have either helicity, and 

\begin{equation}
\Delta u(x) \equiv u_+(x) - u_-(x)  .
\end{equation}

The large $x$ behavior of both $d(x)/u(x)$ and 
$\Delta d(x)/\Delta u(x)$ are of current interest.  Most fits to the
data have the down quarks disappearing relative to the up quarks at
high $x$,  in contrast to pQCD which has definite non-zero predictions
for both of the ratios in the previous sentence.  Recent improved work
on extracting neutron data from deuteron targets, has tended to support
the pQCD predictions~\cite{wally}.  


\begin{figure}

\vskip 5mm

{\epsfxsize 2.4 in \epsfbox{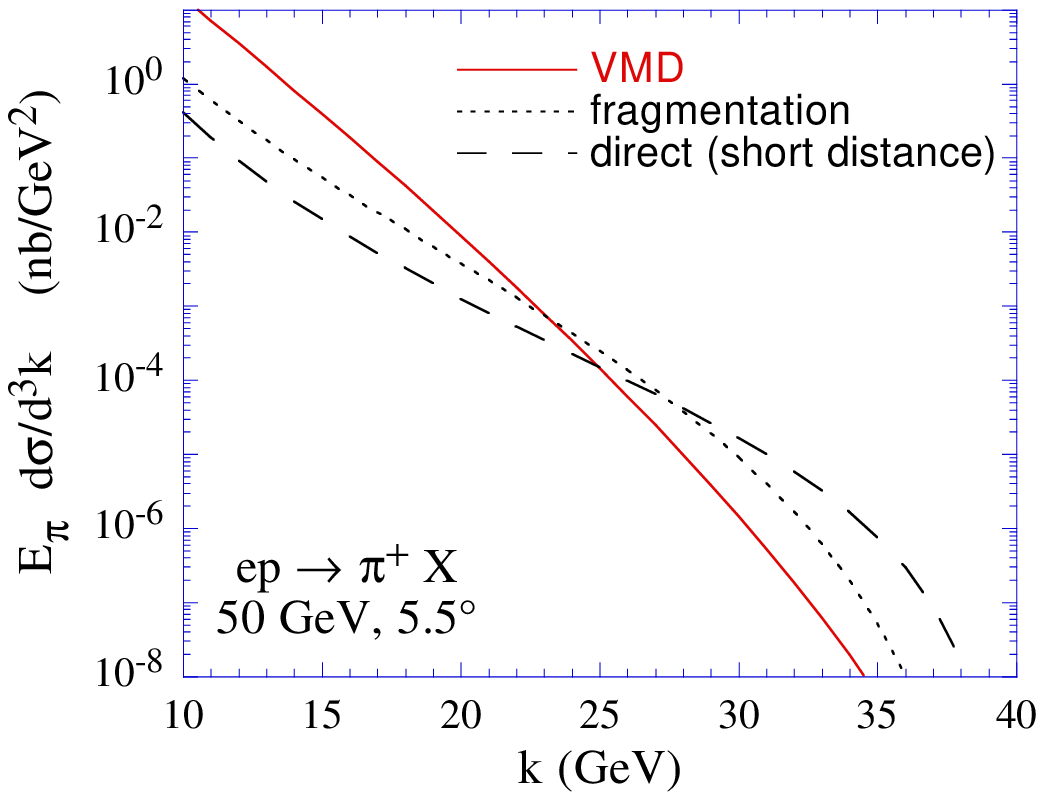}    }

{\vglue -1.84in    \hglue 2.37in
\epsfxsize 2.4 in \epsfbox{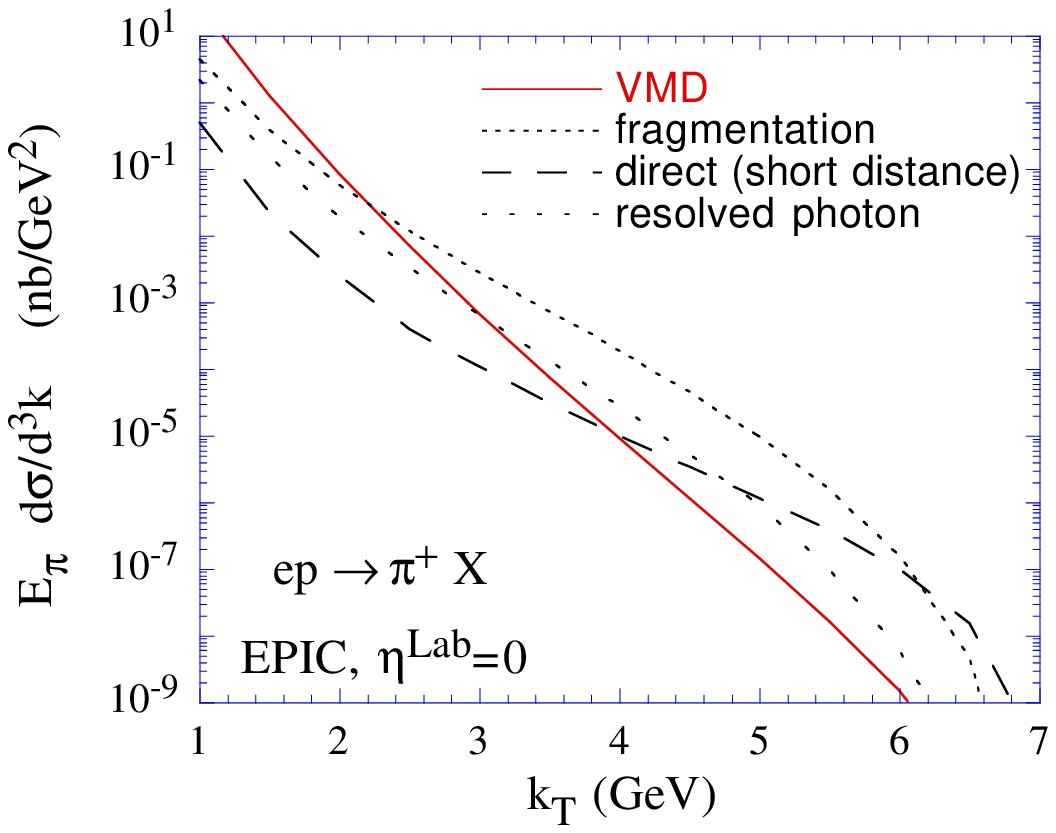}  }

\caption{Calculated contributions to the cross section for 
$ep \rightarrow \pi^+ X$.  Results for SLAC energies with the pion
emerging at 5.5$^\circ$ in the lab are shown on the left. Results for
projected EPIC kinematics, 4 GeV electrons colliding with 40 GeV
protons, with pions emerging at 90$^\circ$ (or rapidity $\eta$ being 0)
in the lab are shown on the right.  The direct process should be quite
important at SLAC (or HERMES), once we have enough transverse momentum
to escape the `soft' region;  EPIC has a long window where the
fragmentation process dominates.}   
\label{sigma}

\end{figure}


There is some data already on $A_{LL}$~\cite{anthony99}, from SLAC End
Station A, which we shall show when we have discussed some of the other
processes that can produce pions.  However, the reader who has gotten
this far should get to see plots that show, at least by calculation,
that there is a non-empty region where direct or short-range pion
production can be seen.  To this end, Fig.~\ref{sigma} shows the
differential cross section for high transverse momentum $\pi^+$
electroproduction for two different kinematics.  The leftmost figure is
for a SLAC energy, 50 GeV incoming electrons, with the pion emerging at
5.5$^\circ$ in the lab. It shows that above about 27 GeV total pion
momentum or 2.6 GeV transverse momentum,  direct (short distance,
isolated) pion production exceeds its competition.   The rightmost plot
is tuned to what I understand is a discussed possibility for EPIC,
namely 4 GeV electrons colliding with 40 GeV protons, with the pions
emerging at 90$^\circ$ in the lab.  Again, the direct pion process
dominates at high enough momentum, although this time the crossover
point is higher and the crossover cross section lower.


\subsection{Moderate $k_T$}

\begin{figure} 

\vskip 5mm

\centerline { \epsfxsize 3in \epsfbox{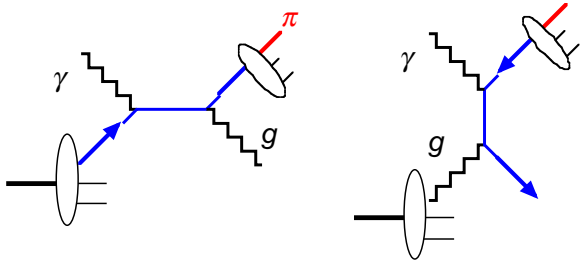}    }

\caption{The fragmentation process.  Pions are produced by
fragmentation of partons at long distances from the primary
interaction region.  The Compton process is on the left; the
pion could come from either quark or gluon fragmentation.  Quark-gluon
fusion is on the right.} 
\label{fragmentation}

\end{figure}

At moderate transverse momentum, the generally dominant process is
still a direct interaction in the sense that the photon interacts
directly with constituents of the target, but the pion is not produced
directly at short range but rather at long distances by fragmentation
of some parton~\cite{many,peralta,acw98}.  Many authors refer to this as
the direct process; others of us are in the habit of calling it the
fragmentation process.  The main subprocesses are called the Compton
process and photon-gluon fusion, and one example of each is shown in
Fig.~\ref{fragmentation}.


\begin{figure} 

\vskip 5mm

\centerline { \epsfxsize 3.4in \epsfbox{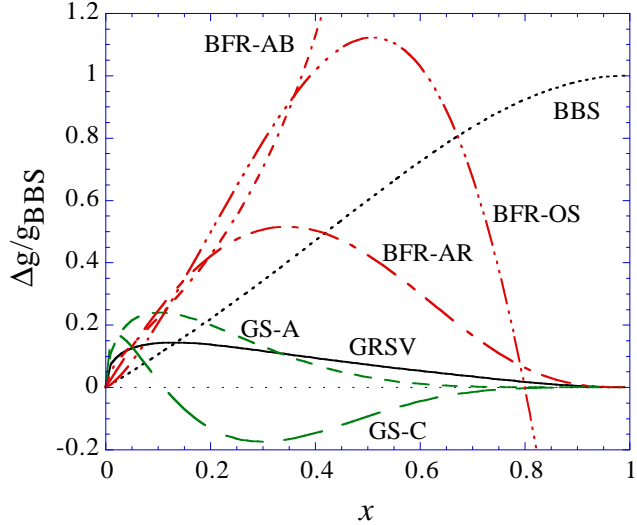}    }

\caption{Possible polarized gluon distributions, all divided by the
same unpolarized distribution.  Most of these are derived from NLO
effects on data and all are claimed to lead to agreement with the
data.  References are BBS~\protect\cite{bbs95};
BFR~\protect\cite{bfr96}; GRSV~\protect\cite{grsv96}; and
GS~\protect\cite{gs96}. } 
\label{gluon7}

\end{figure}

A key feature of hard pion photoproduction in vis the fragmentation
process is that the target gluons are involved in lowest order.   This
stands in contrast to deep inelastic scattering, Drell-Yan processes,
or coincident electroproduction, where target gluons affect the cross
sections only in next-to-leading order (NLO).  These NLO effects can be
significant enough with precise data to give a good determination of
the gluon distribution, and indeed the unpolarized gluon distribution
$g(x)$ has been determined this way.  However, for the polarized
distribution $\Delta g(x)$, the situation is unsettled. 
Fig.~\ref{gluon7} shows a number of different $\Delta g(x)$, normalized
to a common $g(x)$,  that have been derived from analyses of NLO
effects and that all purport to fit the data.  There is clearly need
for additional information, and photon gluon fusion could supply it.
Photon gluon fusion often gives 30--50\% of the cross section for the
fragmentation process, and the polarization asymmetry is as large as
can be in magnitude,

\begin{equation}
\hat A_{LL}(\gamma g \rightarrow q \bar q) = -100\%.
\end{equation}

\noindent Typically for the Compton process, 
$\hat A_{LL}(\gamma q \rightarrow g q) \approx 1/2$.  To use the
fragmentation process we need to have a significant region where that
process dominates, and we need to know the sensitivity of the measured
polarization asymmetry to the different plausible models for $\Delta
g(x)$.  EPIC is the right energy to give a significant region where the
fragmentation process dominates, as may be seen from the right hand part
of Fig.~\ref{sigma}.  The sensitivity is also good, but we shall put
off showing the $A_{LL}$ plots until we discuss the soft processes.


We should also note that the NLO calculations for the fragmentation
process have been done also for the polarized case, though our plots
are based on LO.  For direct pion production,  NLO calculations are
not presently completed.


\subsection{Resolved Photon Processes}

\begin{figure} 

\vskip 5mm

\centerline { \epsfxsize 3.4in \epsfbox{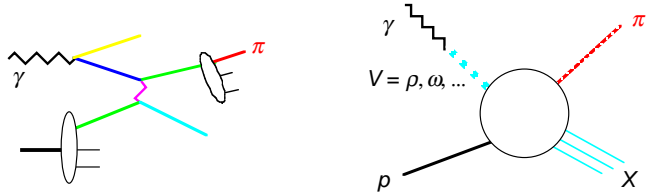}    }

\caption{Resolved photon process (left) and vector meson dominated
process (right).} 
\label{hadronicphoton}

\end{figure}

The photon may split into hadronic matter before interacting with the
target.  If splits into a quark anti-quark pair that are close
together, the splitting can be modeled perturbatively or
quasi-perturbatively, and we call it a resolved photon process. 
Perturbative QCD calculations of the entire process can ensue, and a
typical diagram is shown in the left hand part of
Fig.~\ref{hadronicphoton}. Though resolved photon processes are crucial
at HERA energies, they are never dominant at energies under discussion
here, and we say no more about them.


\subsection{Soft Processes}


This is the totally non-perturbative part of the calculation, whose
size can be estimated by connecting it to hadronic cross sections.  The
photon may turn into hadronic matter, such as 
$\gamma \rightarrow q \bar q + \ldots$ with a wide spatial separation. 
It can be represented as photons turning into vector mesons.  A picture
is shown on the right of Fig.~\ref{hadronicphoton}.

We want a reliable approximation to the non-perturbative cross section
so we can say where perturbative contributions dominate and where they
do not.  Briefly, what we~\cite{acw99} have done to get such an
approximation is to start with the cross section,  given as

\begin{equation}
d\sigma(\gamma A \rightarrow \pi X) = \sum_V {\alpha \over \alpha_V}
            d\sigma(V + A \rightarrow \pi X) + {\rm non-VMD} ,
\end{equation}

\noindent where the sum is over vector mesons $V$, $\alpha = e^2/4\pi$,
and $\alpha_V = f_V^2/4\pi$, with the photon-vector meson vertex in
Fig.~\ref{hadronicphoton} (right side) given as $e m_V^2 / f_V$.  We
can get, for example, $f_\rho$ from the decay 
$\rho \rightarrow e^+e^-$.  

Contributions from the $\rho'$ and other excited $\rho$ mesons are
compensated changing $\alpha_V$ into $\alpha_V^{eff}$, which is about
20\% higher~\cite{ps97}.  Including the $\omega$ and $\phi$ increases
the result by 33\%, according to SU(3).  Now we ``just'' need the cross
section for $\rho^0 A \rightarrow \pi^+ X$.  Lacking direct data, we
approximate it by

\begin{equation}
d\sigma(\rho^0 p \rightarrow \pi^+ X)
   \approx  d\sigma(\pi^+ p \rightarrow \pi^0 X)
   \approx  1.3 d\sigma(\pi^+ p \rightarrow \pi^- X).
\end{equation}
Part of this sequence is backed up by data of O'Neill {\it et
al}~\cite{oneill76}.  Then we used data and data reductions of Bosetti
{\it et al.}~\cite{bosetti73} and Beier {\it et al.}~\cite{beier78} to
get a parameterized fit to the last cross section.  One can compare
what we did to the Regge fit for soft processes of T. Sj\"ostrand {\it
et al.}, which is implemented in PYTHIA~\cite{torbjorn}.

We took the soft processes to be polarization insensitive.  This agrees
with a recent Regge analysis of Manayenkkov~\cite{m99}.


\begin{figure}

\vskip 5mm

{\epsfxsize 2.4 in \epsfbox{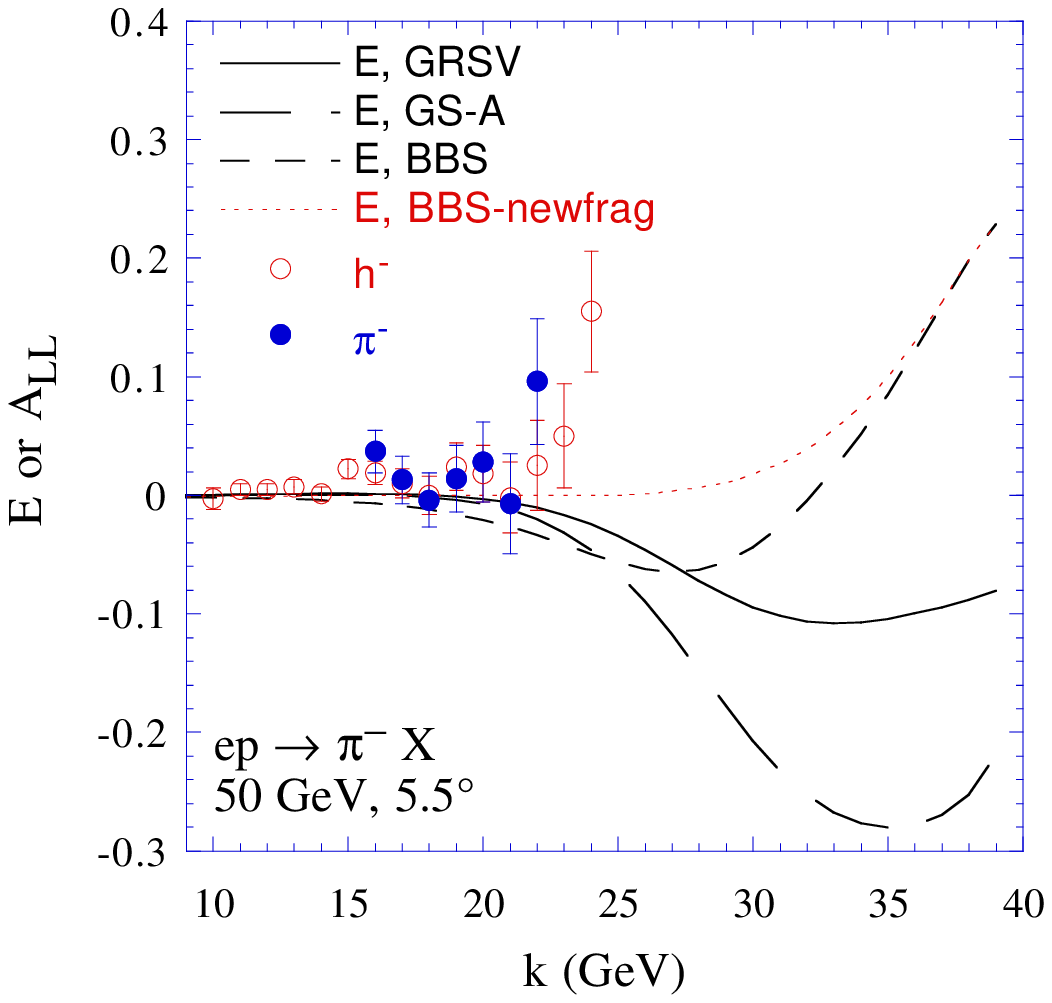}    }

{\vglue -2.26in    \hglue 2.37in
\epsfxsize 2.4 in \epsfbox{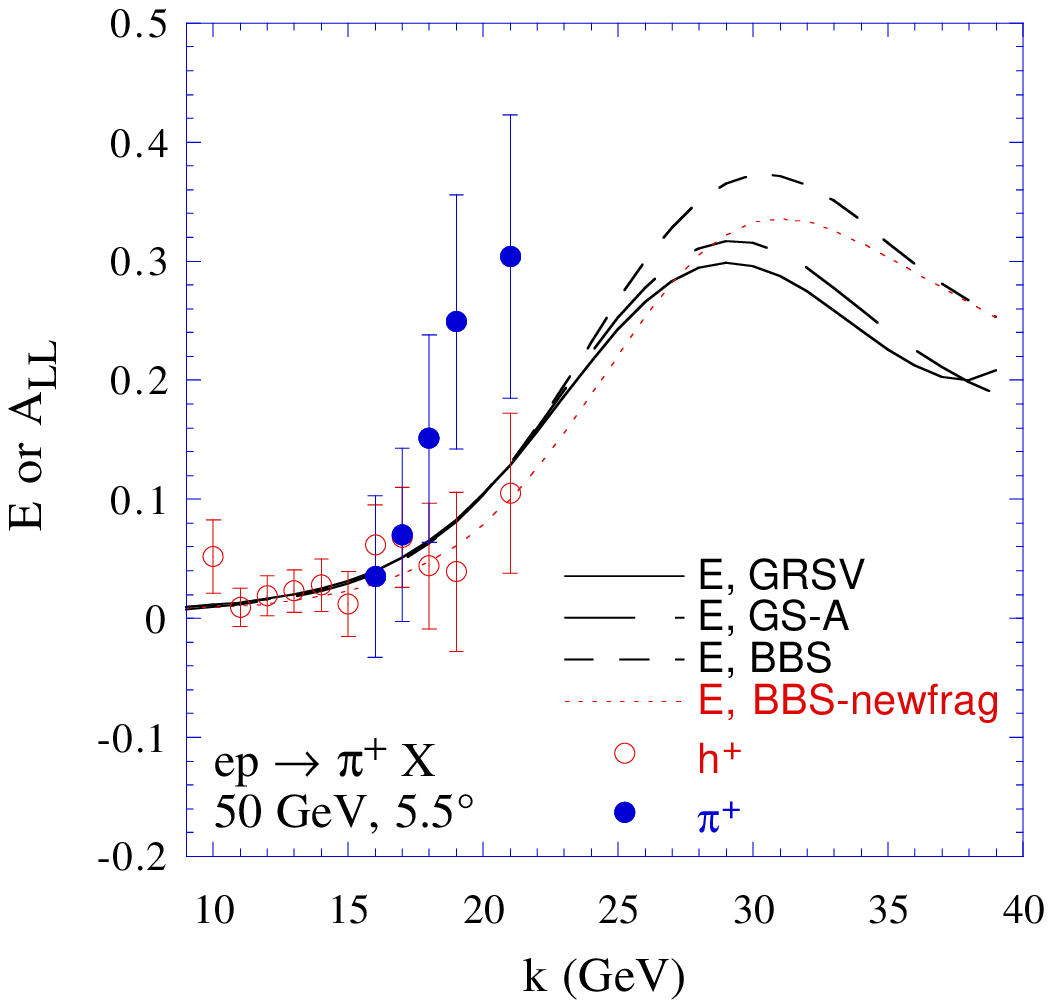}  }

\caption{Longitudinal polarization asymmetries for $\pi^\pm$
production at 5.5$^\circ$ in the lab off proton targets at SLAC
energies.  The calculations are shown for three different polarized
parton distributions~\protect\cite{bbs95,grsv96,gs96} and, for the BBS
case, one non-standard fragmentation function
set~\protect\cite{acw99}.  Much of the data is in the region where the
soft processes are dominant, and the data is consistent with little
polarization sensitivity in the soft region.  The $\pi^+$, however, may
be showing polarization asymmetry from fragmentation or direct pion
production processes coming in at the higher momenta.}   
\label{e_slac_pi_p}

\end{figure}


\section{Results}

Results for the unpolarized cross section have already been displayed
in Fig.~\ref{sigma}.  The soft VMD process is the most important out to
transverse momenta of about 2 GeV.  Above this, at SLAC energies, one
almost immediately enters a region where direct pion production is the
main process.  At planned EPIC energies, however, there is a long
region where the fragmentation process dominates, and this can be of
use is studying $\Delta g$.

Most interesting may be the calculations of $E$ or $A_{LL}$, and
together with the recent data from SLAC.  (Here $E$ is just old notation
for $A_{LL}$.  Barker {\it et al.}~\cite{barker75} in 1975 listed all
measurable asymmetries in pion photoproduction and what we call
$A_{LL}$ was the fifth on their list---and $E$ is the fifth letter of
the alphabet.)  Fig.~\ref{e_slac_pi_p} shows the calculated $A_{LL}$
for both $\pi^-$ and $\pi^+$ off proton targets for three different
parton distribution models.  Although the fragmentation process is not
the crucial one here, we should mention that mostly we used our own
fragmentation functions~\cite{cw93}, and that the results using
BKK~\cite{bkk95} are not very different.  Neither set of
fragmentation functions agrees well with the most recent HERMES
data~\cite{makins} for unfavored vs. favored fragmentation functions,
and the one curve labeled ``newfrag'' is calculated with fragmentation
functions that agree better that data (assuming that data should be
explained by simple fragmentation alone).

Below about 20 GeV total pion momentum, the  soft process dominates and
the data is indeed well described by supposing the soft processes have
no polarization asymmetry.  Above that, the asymmetry is calculated in
pQCD,  and the difference among the results for the different sets of
parton distributions is quite large for the $\pi^-$.

The data of Anthony {\it et al.}~\cite{anthony99} is also shown. 
Presently most, though not all,  of the data is in the region where the
soft processes dominate.  The data is already interesting.  Further
data at even higher pion momenta would be even more interesting,
especially for the
$\pi^-$.  Large momentum corresponds to $x \rightarrow 1$ for the struck
quark,  and pQCD predicts that the quarks are 100\% polarized in this
limit.  Only the parton distributions labeled ``BBS'' are in tune with
the pQCD prediction, and they for large momentum predict even a
different sign for $A_{LL}$ for the $\pi^-$.  The experiment also has
data for deuteron targets, and the calculated results plotted with
the data for this case may be examined in~\cite{acw99}.

\begin{figure}

\vskip 5mm

{\epsfxsize 2.3 in \epsfbox{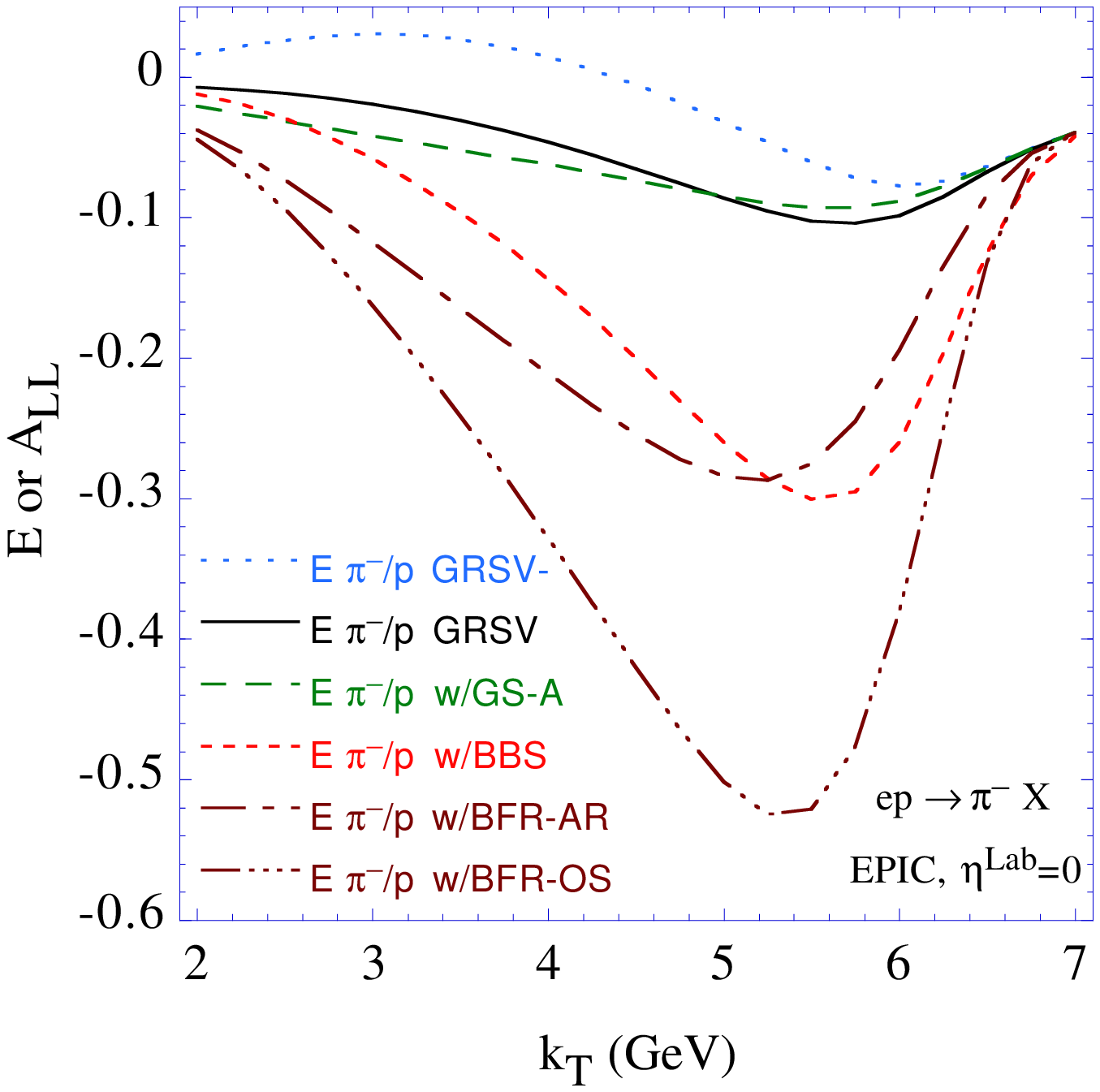}    }

{\vglue -2.25in    \hglue 2.37in
\epsfxsize 2.3 in \epsfbox{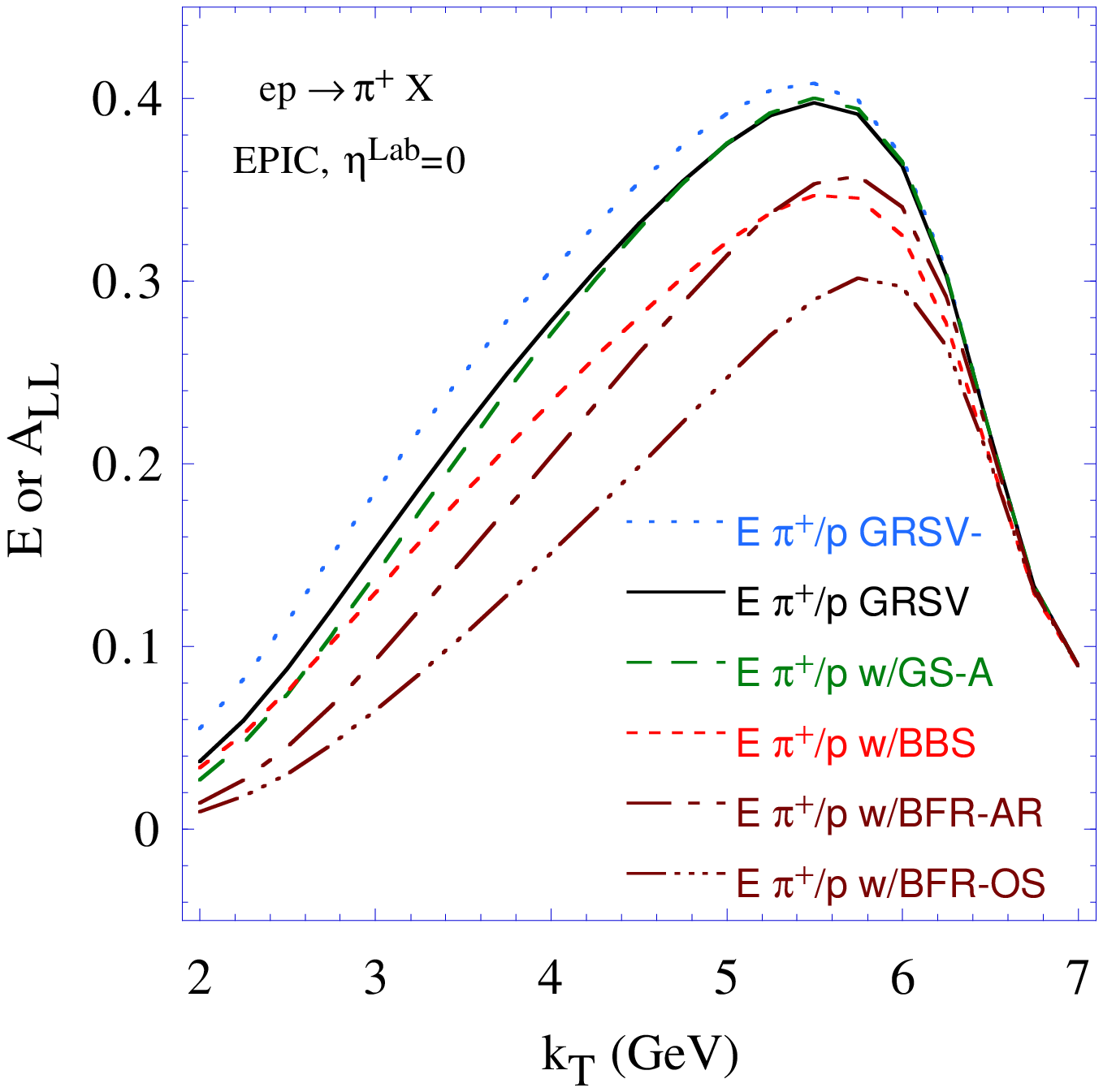}  }

\caption{Polarization asymmetries for different $\Delta g$'s for
projected EPIC energies, with $\pi^\pm$ emerging at 90$^\circ$ or
zero rapidity in the lab.  Each figure has six curves.  All use the
same quark distributions, namely the ones from
GRSV~\protect\cite{grsv96}.  One curve (``GRSV--'') is purely benchmark,
having the polarized gluon distribution
$\Delta g$ set to zero.  The other five curves use five different model
$\Delta g$~\protect\cite{bbs95,bfr96,grsv96,gs96}, as labeled, to show
the potential effectiveness of this measurement for culling gluon
polarization models.}   
\label{epic_pol}

\end{figure}

Regarding EPIC, there is the long region where the fragmentation
process dominates, and we would like to know how sensitive the possible
measurements of $A_{LL}$ are to the different models for $\Delta g$. 
To this end, we present in Fig.~\ref{epic_pol} the results for $A_{LL}$
for one set of quark distributions and 5 different distributions for
$\Delta g$.  The quark distributions and unpolarized gluon
distribution in each case are those of GRSV.  There are 6 curves on each
figure.  One of them is a benchmark, which was calculated with 
$\Delta g$ set to zero.  The other curves use the $\Delta g$ from the
indicated distribution.  There is a fair spread in the results,
especially  for the $\pi^-$ where photon-gluon fusion gives a larger
fraction of the cross section.  Thus, one could adjudicate among the
polarized gluon distribution models.


\section{Summary}                            \label{summary}



Hard, meaning high transverse momentum, semiexclusive processes such as
$\gamma p \rightarrow \pi X$ provide a different way to probe
parton distributions.

There are several perturbative processes that contribute, which we have
called the direct (or isolated or short distance) pion production
process, the fragmentation process, and the resolved photon process. 
All are calculable.  They give us new ways to measure aspects of the
pion wave function, and quark and gluon distributions, especially
$\Delta q$ and $\Delta g$.  The soft processes can be estimated and
avoided if the transverse momentum is greater than about 2 GeV.  EPIC is
projected to have a wide window in which the fragmentation process
dominates, which should make it excellent, in particular, for measuring
$\Delta g$.


\section*{Acknowledgments}
My work on this subject has been done with Andrei Afanasev, Chris
Wahlquist, and A. B. Wakely and I thank them for pleasant
collaborations.  I have also benefited from talking to and reading the
work of many authors and apologize to those I have not explicitly 
cited.  I thank the NSF for support under grants PHY-9600415 and
PHY-9900657.


\end{document}